\input harvmac.tex

\input amssym.def 
\input amssym.tex 

\def\Z{{\Bbb Z}}

\font\huge=cmr10 scaled \magstep2

\def\frac#1#2{{\textstyle{#1\over #2}}}

\def\mod{{\rm mod}}

\def\eqalignD#1{
\vcenter{\openup1\jot\halign{
\hfil$\displaystyle{##}$~&
$\displaystyle{##}$\hfil~&
$\displaystyle{##}$\hfil\cr
#1}}
}

\def\text#1{\quad\hbox{#1}\quad}

\def\lah{{{\lambda}}}\def\muh{{{\mu}}}\def\nuh{{{\nu}}}\def\sih{{\sigma}}
\def\la{\lambda}\def\si{\sigma}\def\om{\omega}

\def\nuh{{\nu}}
\def\muh{{\mu}}

\def\Ra{\Rightarrow}
\def\su{\widehat{su}}
\def\u{\widehat{u}}
\def\osp{\widehat{osp}}

\def\Nc{{\cal N}}

\overfullrule=0pt

\newcount\eqnum  
\eqnum=0
\def\eq{\eqno(\secsym\the\meqno)\global\advance\meqno by1}
\def\eqlabel#1{{\xdef#1{\secsym\the\meqno}}\eq }  

\newwrite\refs 
\def\startreferences{
 \immediate\openout\refs=references
 \immediate\write\refs{\baselineskip=14pt \parindent=16pt \parskip=2pt}
}
\startreferences

\refno=0
\def\aref#1{\global\advance\refno by1
 \immediate\write\refs{\noexpand\item{\the\refno.}#1\hfil\par}}
\def\ref#1{\aref{#1}\the\refno}
\def\refname#1{\xdef#1{\the\refno}}

\def\lad{{\dot \lambda}}
\def\ladd{{\ddot \lambda}}
\def\nud{{\dot \nu}}
\def\nudd{{\ddot \nu}}

\def\mud{{\dot \mu}}
\def\mudd{{\ddot \mu}}

\parskip=6pt

{\nopagenumbers
\ \ \hfill\break 
\vskip1.5cm
\centerline{{\huge Fusion in coset CFT}}
\vskip.25cm  
\centerline{\huge from} 
\vskip.25cm 
\centerline{\huge admissible singular-vector decoupling}
\vskip1.cm
\centerline{ P. Mathieu$^\natural$\foot{pmathieu@phy.ulaval.ca; work supported
by NSERC (Canada) and FCAR (Qu\'ebec).}, 
J. Rasmussen$^\sharp$\foot{rasmussj@cs.uleth.ca; work supported
by NSERC (Canada).} and M.A. 
Walton$^\sharp$\foot{walton@uleth.ca; work supported by NSERC (Canada).}}
\smallskip\centerline{$^\natural$ \it D\'epartement de Physique,
Universit\'e Laval, Qu\'ebec (Qu\'ebec), Canada G1K 7P4}
\smallskip\centerline{$^\sharp$ \it Physics Department, University of
Lethbridge, Lethbridge, Alberta, Canada T1K 3M4}

\vskip1cm
\noindent{\bf Abstract}: Fusion rules for Wess-Zumino-Witten (WZW) models at 
fractional level can be defined in two ways, 
with distinct results. The Verlinde formula
yields fusion coefficients that can be negative. These 
signs cancel in coset fusion rules, however.  
On the other hand, the fusion coefficients calculated from decoupling 
of singular vectors are non-negative. They produce  
incorrect coset fusion rules, however, when factorisation is assumed. 
Here we give two prescriptions that yield the correct coset fusion 
rules from  those found for the WZW models by the
decoupling method. We restrict to the Virasoro 
minimal models for simplicity, and because decoupling results are only
complete in the $\su(2)$ case.

\vfill\break}

\pageno=1

\newsec{Introduction}

The formulation of WZW models in terms of an action is problematic at
fractional admissible level. But their algebraic 
treatment seems to lead to well-defined conformal field theories
(CFTs). 
One vexing point, however, is that the fusion rules do
not seem to be uniquely defined. More precisely, there are two distinct 
ways by which they can be calculated. The Verlinde formula 
[\ref{P. Mathieu and M.A. Walton, { Prog.
Theor. Phys., suppl.} {\bf 102} (1990) 229.}\refname\MW] and the 
decoupling of singular vectors [\ref{H. Awata and Y. Yamada, { Mod.
Phys. Lett.} {\bf A7} (1992) 1185.}\refname\AY] yield different results.  

Although WZW models at fractional level may turn out to be
pathological, certain coset CFTs built from them are consistent
non-unitary models. For example, 
the minimal Virasoro models may be described by the diagonal cosets 
$${\su(2)_k\oplus\su(2)_1\over\su(2)_{k+1}}\ \ .\eqlabel\coco$$
Here $\su(2)_k$ indicates the non-twisted $\su(2)$ Kac-Moody algebra
that is the central extension of the simple Lie algebra
$su(2)$, at fixed level $k$. Now, the fusion rules for the minimal 
models and similar coset theories are
unambiguous. But while the WZW Verlinde fusions have already been
shown to be compatible with them, the decoupling fusions have
not. Here we attempt to fill this gap. We restrict consideration to
the coset (\coco) for simplicity, and because complete results for
decoupling fusion are only known in the $\su(2)$ case.

The fusion rules for the  minimal models $M(p,p')$
take a simple factorised form [\ref{A.A. Belavin, A.M. Polyakov and A.B.
Zamolodchikov, Nucl. Phys. {\bf B241} (1984) {333}.}, 
\ref{{D. Gepner}, Nucl. Phys. {\bf B287} (1987) {111}.}]:
\eqn\fus{\Nc_{(\dot r;\dot s)(\ddot r;\ddot s)}^{[p,p']\quad\,  (r;s)}
= \Nc_{\dot r\,\ddot r}^{[p']\,\; r} \Nc_{\dot s\ddot s}^{[p]\,\;
s}\ \ ,} where $ \Nc_{\dot r\,\ddot r}^{[p']\,\; r}$ 
stands for an $\su(2)$ fusion
coefficient at level $p'-2$. The three labels $\dot r,\ \ddot r$ and $r$ are 
the finite Dynkin labels of three integrable  affine weights, plus one. 
Unitary models are described by the coset (\coco) with
$p=p'+1$ and $p'=k+2$. In the non-unitary case, the same coset 
construction holds, but the level must be fractional [\ref{A. Kent,  {Ph.D.
thesis} (Cambridge University, 1986).}, \MW]: $k=t/u$. Here $u$ is a 
positive integer, while the integer $t$ is relatively prime to $u$ and 
bounded by $t\geq 2-2u$. These
conditions define an admissible level. For the general minimal models, 
unitary and non-unitary both, we have
$$p-p'=u\,,\qquad up'=k+2\ \ .\eq$$
The unitary case is recovered when $u=1$. 

The factorisation \fus\  will be our starting point. We stress that it 
does not rely on special properties of the minimal models other 
than the existence of a diagonal coset
realisation for which one WZW component is at level 1, i.e. (\coco). 
In that case, the integrable weight at the level 1 can be dropped from 
the triplet of weights that together label a coset field. That is because 
the branching rules for the embedding $\su(2)_k\oplus 
\su(2)_1\supset \su(2)_{k+1}$ selecting these 
triplets are very simple. Once the two weights of $\su(2)_k$ 
and $\su(2)_{k+1}$ are fixed, the branching conditions just 
determine the unique level-one weight required as the third label.    
   
In the non-unitary case, the fractional part of the branching
condition leads to a simple constraint: the finite (horizontal) 
parts of the fractional parts
of the two non-integrable weights must be equal. If this is satisfied,
the original (though slightly modified)
branching condition plays the same role as in the unitary
cosets. It simply determines the level-one weight required in the
triplet of weights labelling the coset field.    

In non-unitary coset models, there is a large number of 
field identifications. These include 
types not present in their unitary analogues. It has been
proved in [\ref{P. Mathieu, D. S\'en\'echal and M.A. Walton, {
Int. J. Mod. 
Phys.} {\bf A7} (1992) 731.}\refname\MSW, \ref{P. Mathieu and
M.A. Walton, Nucl. Phys. {\bf B553}
(1999) 533.}\refname\MWa] that we can always choose a coset-field 
representative whose three weights
have vanishing finite fractional parts.  For fusions involving such  
weights, the two methods  yield identical results. However, the coset 
fusion rules should be computable with any choice of 
coset representatives,  and the result must be independent of this choice. 
It is  natural then, to test the two methods of calculating 
admissible fusions by using coset
representatives with non-vanishing finite fractional parts.

The Verlinde formula generates negative WZW fusion coefficients for 
WZW models at fractional admissible levels [\MW, \ref{I.G. Koh and
P. Sorba, {Phys. Lett.} {\bf B215} (1988) 723.}], perhaps indicating  
an ill-defined theory.  In coset models, however, the Verlinde method 
always leads to positive fusion coefficients as all the negative signs
cancel. Moreover, for the coset (\coco), it reproduces \fus.  

On the other hand, singular-vector 
decoupling leads to non-negative fusion coefficients.  From that point of 
view then, the non-unitary WZW models look well-defined. 
The naive application of these fusion coefficients to the computation 
of coset fusion rules seems to fail, however: assuming factorisation, 
even the multiplicities are not correct. 

But the relation between the WZW fusions and the coset ones may need to be 
modified in the fractional case. Here we search for that modification, and 
find a new product of component fields that yields results compatible with 
the Verlinde method. It is 
evocative (in being formulated as a sort of supertrace) 
of the $\osp(1|2)$
Lie  superalgebra pattern of the $\su(2)$ 
singular-vector decoupling fusions, that has  
already been described in [\ref{B. Feigen and F. Malikov, Modular
functor and representation theory of $\hat{sl}_2$  
at fractional level, 
q-alg/9511011.}\refname\FM].

Alternatively, the coset fusion rules may be obtained by using certain
truncations of the
singular-vector decoupling fusion. 
Several possibilities exist. We discuss some of them, and we 
provide a natural motivation for a particular truncation, while relating it 
to Verlinde fusion. 

\newsec{Admissible representations of $\su(2)_k$}

The set of admissible representations contains the set of 
integrable representations. At fixed level, the characters of admissible 
representations form a finite-dimensional representation of the modular 
group [\ref{V. Kac and M. Wakimoto, { Proc. Nat. Acad. Sci. USA} {\bf 85}
(1988) 4956, { Adv. Ser. Math. Phys.}
{\bf 7} (World Scientific, 1988) p.138.}\refname\KW].  
The admissible $\su(2)_k$ representations are the building 
blocks in the algebraic formulation of the WZW models at admissible
level $k$, and are in one-to-one correspondence with the WZW primary fields.

Admissible representations of $\su(2)_k$ at fractional admissible level 
$k=t/u$ can be described rather simply. The highest weight 
$\lah$ of such a representation can be written 
as
\eqn\IF{\lah = \lah^I -(k+2)\lah^F\ ,}
in terms of two integrable weights $\lah^I$ and $\lah^F$   
at respective levels
\eqn\kIkF{\eqalign{
k^I =~& u(k+2) -2 \geq  0\ ,\cr
k^F =~& u-1 \geq 0\ .\cr
}}    
The superscripts $I$ and $F$
refer to integer and fractional. Notice that although 
$\lah^F$ is responsible for the fractional part of $\lah$, it is itself 
an integrable weight.  We will  use the Dynkin label description
$$\lah = \la_0{{\omega}}_0+\la_1{{\omega}}_1\ ,\eq$$
where the ${{\omega}}_i$ are the affine fundamental weights.\foot{Since in
the present work only affine weights are used, the usual ``hats'' 
are omitted for simplicity.} An affine weight $\lambda$ at level $k$
has 
\eqn\levk{\lambda_0 + \lambda_1\ =\ k\ ,}
so that if $k$ is fixed, $\lambda_1$ specifies the weight
uniquely. Consequently, $(\la^I_1, \la^F_1)$ specifies an admissible
$\su(2)_k$ weight, and this shorthand notation will prove useful in the
following.

\newsec{WZW fusion rules from the Verlinde formula}

The Verlinde formula leads to the following expression for the 
$\su(2)_k$ fusion rules at fractional level $k$ [\MW]:
$$\lah\times \muh\ \  = \sum_{\nu^I_1\in \delta_{k^I}\atop
\nu^F_1\equiv\la^F_1+\mu^F_1~(\mod~ u)} ~(-a^I)^{[(\la^F_1+\mu^F_1)/ u
]}~\nuh\ .\eqlabel\frafus$$
Here the set $\delta_{k^I}$ is defined as
$$
\delta_{k^I}= \left\{\, \nu^I_1\ |\ \la^I_1+\mu^I_1 + 
\nu^I_1\equiv 0~ (\mod~2);\quad
|\la^I_1-\mu^I_1|\leq \nu^I_1 \leq k^I-|k^I-\la^I_1 -
\mu^I_1|\,\right\}\ ,
\eqlabel\sett
$$
and the square brackets denote the integer part:
\eqn\intp{ 
 [(\la^F_1+\mu^F_1)/u]\ =\  \left\{\matrix{
 0\qquad &{\rm if}\quad \la^F_1+\mu^F_1<u\cr
 1\qquad &{\rm if}\quad \la^F_1+\mu^F_1\geq u\cr}\right.\ \ .
} 
$a^I$ is the outer automorphism with its  action restricted to the 
integer part of $\nuh$:
$$ a^I \nuh = a\nuh^I -(k+2)\nuh^F\ .\eq$$
$a$ itself simply interchanges the two Dynkin labels of an affine 
weight:  
$$a\lambda= a(\lambda_0\omega_0 +\lambda_1\omega_1)\ =\ \lambda_1\omega_0
 + \lambda_0\omega_1\ =\ k(\omega_0+\omega_1) - \lambda\ \ ,\eq$$
where \levk\ was used in the last equality. Of course,
\eqn\aIF{a\la\ =\ a\left(\la^I - (k+2)\la^F\right)\ =\ a\la^I -
(k+2)a\la^F\ .} In convenient notation,
$$ 
 a(\la^I_1,\la^F_1)=(k^I-\la^I_1,k^F-\la^F_1)\ \ .
$$

The result (\frafus) is proved  along the following lines.
We first relate the ratio 
$\gamma_{\lah}^{(\sih)} = {S_{\lah, \sih} / S_{0,\sih}}$
to the finite character $\chi_\la$ evaluated at $\xi_{\si}= -2\pi i 
{(\si+\rho)/( k+2)}$.
The line of the argument is then standard: the $\gamma_{\lah}^{(\sih)}$'s 
satisfy the fusion rules, while the product
of $\chi_\la (\xi_{\si})$'s is decomposed using the 
tensor-product coefficients. The fusion coefficients at level
$k$ are thus related to tensor-product coefficients and then to 
fusion coefficients at level
$k^I$.\foot{That it is $k$ and not $k^I$  that appears in the sine factor of
the $S$ matrix prevents a more direct relation between 
$\Nc^{(k)}$ and $\Nc^{(k^I)}$.} When $\la^F_1+\mu^F_1<u$, the above result
is immediate. In that case the
fusion coefficients are non-negative, and reduce to the
$\su(2)_{k^I}$ fusion coefficients with a $\u(1)$ factor arising 
from the fractional parts (the $\u(1)$ interpretation means that the 
fractional parts simply add ($\mod~u$)). When
$\la^F_1+\mu^F_1\geq u$, we can write
$$\la^F+\mu^F = \nu^F + u\zeta\ \ .\eq$$
Clearly, $\zeta$ can always be written in the form $\zeta_1a\om_0$.
This $\zeta$ piece is responsible for an extra phase factor. 
Up to a sign, this phase factor can be absorbed in the transformation
$\chi_{\nu^I}\rightarrow \chi_{a \nu^I}$.  The resulting minus
sign is the one that makes fusion coefficients negative when  
$\la^F_1+\mu^F_1\geq u$. The fusion coefficients are now expressed
in terms of $\Nc_{\lah^I \muh^I}^{~~~a\nuh^I}$.  

Ignoring outer automorphisms and minus signs, the pattern of fusion 
rules as computed with the Verlinde formula is 
$$\Nc[\su(2)_k]\sim \Nc[\su(2)_{k^I}]\; \Nc[\u(1)_{k^F}]\ .\eqlabel\verp$$
In the last factor we have indicated that the $\u(1)$ may be 
interpreted to have level $k^F=u-1$. 

We should stress that the computations of fusion rules via  
BRST cohomology  [\ref{D.
Bernard and G. Felder, { Commun. Math. Phys.} {\bf 127} (1990) 145.}]
or vertex-operator  methods [\ref{C. Dong, H. Li and 
G. Mason, { Commun. Math. Phys.} {\bf 184} (1997) 65.}] 
agree with those just written when $\la^F_1+\mu^F_1< u$.

\newsec{Coset fusion rules from the Verlinde formula}

Now let $\lah$ and $\nuh$ stand for the admissible weights at levels   
$k$ and $k+1$, respectively, that label a coset
field $\{\lah,\nuh\}$. Let two other coset fields be labeled 
$\{\dot\lah,\dot\nuh\}$, $\{\ddot\lah,\ddot\nuh\}$ in a similar way. 
The coset fusion rules take the form
\eqn\cosf{
\Nc_{\{\dot\lah,\dot\nuh\}\{\ddot\lah,\ddot\nuh\}}^{\qquad
\{\lah,\nuh\}}\ =\  
\Nc_{\dot\lah{\ddot\lah}}^{(k)\,\; {\lah}}
\Nc_{\dot\nuh{\ddot\nuh}}^{(k+1)\,\; {\nuh}}\ .}
Using (\frafus), this becomes 
\eqn\cosfi{ 
\Nc_{\{\dot\lah,\dot\nuh\}\{\ddot\lah,\ddot\nuh\}}^{\qquad
\{\lah,\nuh\}}\ =\  \left\{\matrix{
\Nc_{\dot\lah^I{\ddot\lah^I}}^{(k^I)\quad {\lah^I}}
\Nc_{\dot\nuh^I{\ddot\nuh^I}}^{(k^I+u)\,\; 
{\nuh^I}}\qquad &{\rm if}\quad
\dot\lah^F_1+\ddot\lah^F_1 <u\cr  
\Nc_{\dot\lah^I,{\ddot\lah^I}}^{(k^I)\,\; {a\lah^I}}
\Nc_{\dot\nuh^I,{\ddot\nuh^I}}^{(k^I+u)\,\; 
{a\nuh^I}}\qquad &\ \ {\rm if}\quad 
\dot\lah^F_1+{\ddot\lah^F_1}
\geq u\ \ .\cr}\right.
}
Notice that if $\dot\lah^F_1+\ddot\lah^F_1<u$, it also follows that
$\dot\nuh^F_1+\ddot\nuh^F_1 <u$ since $\lah$ 
and $\nuh$ are part of the same coset field, and hence have the 
same fractional part. However, the coset field identification
$$\{\lah,\nuh\} = \{a\lah, a\nuh\}\eqlabel\fid$$
holds in the non-unitary case as well as in the unitary one. Therefore,  
the two expressions in \cosfi\ are equivalent. The result is equivalent 
to \fus; the latter is expressed in terms of $\rho$-shifted weights. This 
result is manifestly independent of the choice of
the coset-field representative since here we have made no choice of 
the value of the fractional part.  

\newsec{WZW fusion rules from singular-vector decoupling}

Fusion rules can be calculated by enforcing 
the decoupling of singular vectors. In the integrable WZW  case 
[\ref{A.B. Zamolodchikov and V.A. Fateev, Sov. J. Nucl. Phys. 
{\bf 43} (1986) 657.}], this method leads to results that agree 
with those obtained by the Verlinde formula.
Fractional levels present the novelty of {\it two} types 
of solutions, called A and B below. The admissible representations 
$\nuh$ that appear in the product $\lah\times\muh$ are 
[\AY]\foot{The parameters in [\AY] are related to ours as follows:
$r_{\rm AY}= \la^I_1+1$ and $s_{\rm AY}=\la_1^F$, $t_{\rm AY}= k+2$, 
$p_{\rm AY}= t+2u$ and
$q_{\rm AY}=u$.}:
$$\eqalignD{ 
&{\rm Case~ A:}\quad &|\la^I_1-\mu^I_1 |\leq 
 \nu^I_1\leq \; k^I-|k^I-\la^I_1 -\mu^I_1| \cr
&~ &|\la_1^F-\mu_1^F |\leq  
\nu_1^F\leq \; k^F-|k^F-\la^F_1 -\mu^F_1|\ \ , \cr
&~&\cr 
&{\rm Case~ B:}\quad &|\la^I_1-\mu^I_1 |\leq 
 k^I-\nu^I_1\leq \; k^I-|k^I-\la^I_1 -\mu^I_1| \cr
&~ &|\la_1^F-\mu_1^F |+1\leq 
k^F - \nu_1^F\leq \; k^F-1 - |k^F-\la^F_1 -\mu^F_1|\ \ . \cr}\eqlabel\ayf$$

\noindent In all 4 lines just written, the bounded quantities take all values
from their lower bounds to their upper bounds, increasing in steps of 2. 

For later use, we separate the fusion coefficients determined by
(\ayf) into A and B parts:
\eqn\nab{{\cal N}_{\la\mu}^\nu\ =\ {\cal A}_{\la\mu}^\nu\ +\ 
{\cal B}_{\la\mu}^\nu\ \ . } Notice that a B-type solution is intrinsically
fractional, that is, ${\cal B}_{\la\mu}^\nu=0$ when $u=1$. More
precisely, the B-set of solutions can be non-empty
only if the following bounds are satisfied:
$$1\leq \la_1^F, \mu_1^F, \nu_1^F \leq u-2\ \ .\eq$$
Hence, not only do  the fractional parts need to be non-zero, 
but $u$ must be greater than 2.
 
Case A describes an $\su(2)_{k^I} \times
\su(2)_{k^F}$ fusion rule, i.e., a separate $\su(2)$ fusion
for each of the integer and the fractional parts:
\eqn\ANN{
 {\cal A}_{\la\mu}^\nu\ =\ {\cal N}_{\la^I\mu^I}^{(k^I)\,\; \nu^I}
 {\cal N}_{\la^F\mu^F}^{(k^F)\,\; \nu^F}\ \ .
}
Similarly, case B may be factorised into two $\su(2)$ fusions, though
with shifted weights and level in the fractional sector: 
\eqn\BNN{
 {\cal B}_{\la\mu}^\nu\ =\ {\cal N}_{\la^I\mu^I}^{(k^I)\,\; a\nu^I}
 {\cal N}_{\la^F-\omega_0-\omega_1\,\ \ \mu^F-\omega_0-\omega_1}^{
  (k^F-2)\,\; a(\nu^F-\omega_0-\omega_1)}\ \ .
}

A further observation is that the combined set of A and B fusions
(\ayf) displays an $\osp(1|2)_k$ fusion pattern. We
recall that $\osp(1|2)_k$ fusion rules are equivalent to those for 
$\su(2)_k$, except that the constraint
$\la_1+\mu_1-\nu_1\equiv 0$ (mod 2) is weakened to 
$\la_1+\mu_1-\nu_1\equiv 0$ (mod 1) [\ref{J-B. Fan and M. Yu,
{\it Modules over affine Lie superalgebras}, hep-th/9304122;
I.P. Ennes, A.V. Ramallo and J.M. Sanchez de Santos, Nucl. Phys. {\bf
B491} (1997) 574.}\refname\ospfr]. That is, all integer intermediate
values of the Dynkin label $\nu_1$ are allowed.
The fusion rules obtained by the singular-vector decoupling method 
are seen to allow sets of weights isomorphic to those allowed by
$\osp(1|2)$ fusion rules. It is the shifts in the lower and upper
bounds in case B of (\ayf) that allow this simple interpretation. 
Thus the case-B solutions correspond to fermionic contributions in this
analogy. 

Let us inspect further the set of weights appearing in the 
fusion rule encoded in (\ayf). For a fixed non-vanishing
fusion $\lambda\times \mu$, there is always
one more fractional A-type solution $\nu^F$ than fractional B-type solution.
Furthermore, as already indicated in \ANN\ and \BNN, each of the four
lines in (\ayf) corresponds to an $\su(2)$ fusion. Finally, in order
to emphasise the similarity between the two types, one is written
in terms of $\nu$ while the other is written in terms of $a\nu$.
We also note that acting with $ac$ on a weight with non-vanishing
fractional part has the effect of reducing the fractional part
(of the first Dynkin label) by one while leaving the integer part
unchanged. Here $c$ represents the action of charge conjugation
as defined by the square of the modular matrix $S$: $(S^2)_{\lambda,\mu} = 
\pm \delta_{c\lambda,\mu}$. Explicitly, we have
\eqn\c{
 c(\la^I_1,\la^F_1)=
 \delta_{\la^F_1,0}(\la^I_1,\la^F_1)+(1-\delta_{\la^F_1,0})(k^I
 -\la^I_1,u-\la^F_1)\ \ . 
}
In conclusion, for fixed $\nu^I$, the fractional parts paired with
it form the pattern illustrated by the following diagram: 
\eqn\diag{\eqalign{
 &\cr
 {\cal A}_{\la\mu}^{(\nu^I_1,k^F-|k^F-\la^F_1-\mu^F_1|)}&\cr
 \qquad&c\searrow\cr
 \qquad&\qquad
  {\cal B}_{\la\mu}^{a(\nu^I_1,k^F-|k^F-\la^F_1-\mu^F_1|-1)}\cr
 \qquad&\swarrow {c}\cr
 {\cal A}_{\la\mu}^{(\nu^I_1,k^F-|k^F-\la^F_1-\mu^F_1|-2)}&\cr
 \qquad&c\searrow\cr
 \qquad&\ \vdots\cr
 \qquad&c\searrow \cr
 \qquad&\qquad
  {\cal B}_{\la\mu}^{a(\nu^I_1,|\la^F_1-\mu^F_1|+1)}\cr
 \qquad&\swarrow{c}\cr
 {\cal A}_{\la\mu}^{(\nu^I_1,|\la^F_1-\mu^F_1|)}&\ .\cr &\cr
}}
All arrows indicate an action of the charge conjugation $c$ 
on the upper weight. All fusion 
coefficients ${\cal A}$ and ${\cal B}$ depicted here have identical
value, i.e. 0 or 1. 

Thus, up to the action of the outer 
automorphism group, we have the following schematic structure
$$\Nc[\su(2)_k]\sim\Nc[\su(2)_{k^I}]\; \Nc[\osp(1|2)_{k^F}]\ .\eq$$
This is very different from the pattern that follows 
from the Verlinde formula (\verp). The occurrence of the $\osp(1|2)$ 
pattern makes contact with the results obtained in [\FM].

We would like to stress that the result (\ayf) obtained in 
[\AY] is supported by 
calculations of four-point functions using a free-field 
realisation [\ref{J.L. Petersen, J. Rasmussen and M. Yu, { Nucl. Phys.} 
{\bf B457} (1995) 309, { Nucl. Phys.} {\bf B481} (1996) 577;
J. Rasmussen, Ph.D. thesis (Niels Bohr Institute, 1996)
hep-th/9610167.}\refname\PRYR]
(see also [\ref{P. Furlan, A.Ch. Ganchev and V.B. Petkova, { Nucl. Phys}.
{\bf B491} (1997) 635.}] for similar results from  a somewhat 
different approach). These correlation functions satisfy 
the Knizhnik-Zamolodchikov equations and are invariant under projective
and duality transformations, 
testifying for their soundness. Generalisations are reported in 
[\ref{J.L. Petersen, J. Rasmussen and M. Yu, { Nucl. Phys.} {\bf B502}
(1997) 649; J. Rasmussen, { Mod. Phys. Lett.} {\bf A13}
(1998) 1281, Int. J. Mod. Phys. {\bf A14} (1999) 1225.}]. 

\newsec{WZW fusion rules: comparing the two methods}

To compare the results of the two methods at the WZW level, 
consider the special case $k=-4/3$ ($k^I=0$ and $ k^F=2$). There are 
$(k^I+1)(k^F+1)=3$ admissible representations. In the notation
$(\la_1^I, \la_1^F)$, these are $(0,0),\; (0,1)$ and $(0,2)$.
The Verlinde formula produces the fusion rules
$$\eqalign{ 
& (0,1)\times (0,1) = \phantom{-}(0,2)\ \ ,\cr
&(0,1)\times (0,2) = -(0,0)\ \ ,\cr
&(0,2)\times (0,2) = -(0,1)\ \ .\cr}\eqlabel\exv$$
We see the minus signs arising in the cases where $\la_1^F+\mu_1^F\geq u=3$.
On the other hand, the results of [\AY] give (combining solutions 
from cases A and B, cf. (\ayf)):
$$\eqalign{ 
&(0,1)\times (0,1) = (0,0)+(0,1)+ (0,2)\ \ ,\cr
&(0,1)\times (0,2) = (0,1)\ \ ,\cr
&(0,2)\times (0,2) = (0,0)\ \ .\cr}\eqlabel\exd$$
Case B leads to a non-zero contribution only  for the first fusion: 
this is the $(0,1)$ representation. We see the $\osp(1|2)$-like
structure showing up in the fractional sector: in the first
example, there appears an intermediate weight between those 
predicted by the $\su(2)_2$ fusion rule. 
Let us emphasise that since $k^I=0$ and $k^F=2$, $a(0,1)=(k^I-0,k^F-1)=(0,1)$.

One can easily verify that the fusion matrices defined by (\exd) 
are not diagonalisable by any symmetric matrix. On the other hand, the
Verlinde formula guarantees that the symmetric modular-$S$ matrix 
diagonalises the fusion matrices encoded in (\exv). 

This simple example illustrates important 
differences between the two methods of computing the WZW fusion rules. 
Due to the lack of a physical realisation of WZW models at 
fractional level, it seems difficult to favour one method over the other. 
However, given that the number of weights appearing
in the fusion is quite different in the two cases, one 
might expect notable differences in the coset models.
 
\newsec{Coset fusion rules from WZW singular-vector decoupling}

Let us consider the simplest non-unitary minimal model, the 
Yang-Lee singularity. It has $(p,p') = (5,2)$, and hence $k=-4/3$ in the 
diagonal coset description. There are 12 coset fields that 
can be grouped into two sets among which all 6 fields are
identified. That is, there are two equivalence classes of coset 
labels under field identification, corresponding to the two Yang-Lee
primary fields (see [\MSW], reviewed in sect. 18.7.2 of [\ref{P. Di
Francesco, P. Mathieu and
D. S\'en\'echal,  {\it Conformal field theory} 
(Springer, 1997).}\refname\DMS], 
for details about the field 
identifications, and [\MWa] for an improved approach).   
In the notation $\{(\la_1^I, \la_1^F), \mu_1; (\nu_1^I,\nu_1^F)\}$, 
where the three weights are at
levels $k=-4/3,\,1$ and $k+1=-1/3$, respectively, they are:
\eqn\list{\eqalign{
h=0\qquad  & \{(0,0),0;(0,0)\} \quad  \{(0,2),1;(3,2)\} 
  \quad\{(0,1),1;(0,1)\}  \cr
 &\{(0,1),0;(3,1)\} \quad  \{(0,2),0;(0,2)\} \quad \{(0,0),1;(3,0)\}\ \ ,\cr
h=-\frac15\qquad & \{(0,0),1;(1,0)\} \quad  \{(0,2),0;(2,2)\} 
 \quad  \{(0,1),0;(1,1)\}
 \cr & \{(0,1),1;(2,1)\} \quad  \{(0,2),1;(1,2)\} 
 \quad  \{(0,0),0;(2,0)\}\ \ .\cr}}  
For our example, we will focus on the $h=0$ identity field.

Let us pick $\{(0,1),1;(0,1)\}$ and $\{(0,1),0;(3,1)\}$ as
representatives of the identity field, 
and omit the level-one weights, relabelling our choices  
as $\{(0,1);(0,1)\}$ and $\{(0,1);(3,1)\}$. We calculate the fusion
rule of the identity field ($I\times I= I$) using these
representatives, computing the level
$k$ and the level $k+1$ fusions independently:
$$\eqalign{
\{(0,1);(0,1)\}\times \{(0,1);(3,1)\} &= 
\{ (0,1)\times (0,1);(0,1)\times (3,1)\}\cr
 &= \{ (0,0)_{\rm A}+(0,1)_{\rm B}+(0,2)_{\rm A} ; 
  (3,0)_{\rm A}+(0,1)_{\rm B}+(3,2)_{\rm A}\}\ \ .\cr}\eq$$
The subscripts indicate case A and B results. Next we combine the
separate WZW  fields 
into coset fields. A simple product of the
different fields leads to 9 possibilities. Not all resulting  
combinations respect the coset
branching rules, however. If we restrict ourselves to the ones that
do, we get only three candidates:
\eqn\thr{
\{(0,0)_{\rm A}; (3,0)_{\rm A}\} + \{(0,1)_{\rm B}; (0,1)_{\rm B}\}
 +\{(0,2)_{\rm A}; (3,2)_{\rm A}\}\ \ .
}
All of these three fields are representatives of the identity, as  
they should be (cf. the list \list). However, the 
fusion coefficient of the identity with itself is then 3 rather than 1. 

A facetious cure for that discrepancy is to introduce a minus sign in 
front of the second coset field, transforming $1+1+1$ into $1-1+1$.  
But this illustrates our general proposal. We show that (i) only
fusions that are of type A, or type B, in {\it both} labels should
contribute, and suggest that (ii) the type B contributions should be
negative. Because of property (i), the proposal is of a trace
form. As already mentioned, case B is analogous to fermionic
contributions in $\widehat{osp}(1|2)$ fusion. The minus sign of
property (ii) thus gives our proposal a structure similar to a
supertrace. 

\newsec{A coset ``supertrace''}

In this section, we will describe our general proposal explicitly, and
prove it computes the correct  minimal model fusion rules. 

Consider the fusion of the two coset fields:
\eqn\cos{\eqalign{
 \big\{(\lad^I_1,\lad^F_1);(\nud^I_1,\nud^F_1)\big\}\ \ ,
  \ \ \ \ &\lad^F_1=\nud^F_1\ \ ,\cr
 \big\{(\ladd^I_1,\ladd^F_1);(\nudd^I_1,\nudd^F_1)\big\}\ \ ,
  \ \ \ \ &\ladd^F_1=\nudd^F_1\ \ .\cr 
}}
Our main result is the following proposal:
\eqn\str{\eqalign{
\big\{(\lad_1^I&,\lad_1^F);(\nud_1^I,\nud_1^F)\big\}\times
 \big\{(\ladd^I_1,\ladd^F_1);(\nudd^I_1,\nudd^F_1)\big\}\cr
&\ =\ {\rm str}\big\{(\lad^I_1,\lad^F_1)\times(\ladd^I_1,\ladd^F_1);
 (\nud^I_1,\nud^F_1)\times(\nudd^I_1,\nudd^F_1)\big\}\cr
&\ =\ \sum_\la\sum_\nu\left({\cal A}_{\lad\ladd}^\la{\cal A}_{\nud\nudd}^\nu
 -{\cal B}_{\lad\ladd}^\la{\cal B}_{\nud\nudd}^\nu\right)
 \delta_{\la^F_1,\nu^F_1}\big\{(\la^I_1,\la^F_1);(\nu^I_1,\nu^F_1)\big\}
 \ \ .\cr 
}}
The delta function is due to the branching conditions. It is the only 
condition to impose here as we do not specify the weight $\mu$.  
The latter is given by the remaining branching condition
\eqn\branch{\mu_1\equiv\nu^I_1-\la^I_1-\la^F_1\ \ (\mod~2)\ \ .}
Similar branching conditions apply to the coset fields \cos. 
The finite double-summation in \str\ may be written
\eqn\strii{
\sum_{\la^F}\sum_{\la^I,\nu^I}\left({\cal A}_{\lad\ladd}^{(\la^I,\la^F)}
 {\cal A}_{\nud\nudd}^{(\nu^I,\la^F)}
 -{\cal B}_{\lad\ladd}^{(\la^I,\la^F)}{\cal B}_{\nud\nudd}^{(\nu^I,\la^F)}
 \right)\big\{(\la^I_1,\la^F_1);(\nu^I_1,\la^F_1)\big\}\ \ .
}
Our proof that this reduces to \fus\ relies on the structure of field
identifications [\MSW, \MWa]. Those that are important here can be
written in terms of the outer (diagram) automorphism $a$, and the
operation $c$, related to charge conjugation, as defined by
$S^2$. Their actions on (first Dynkin labels of) coset weights are  
\eqn\accoset{\eqalign{
a\big\{(\la^I_1,\la^F_1);(\nu^I_1,\la^F_1)\big\}=&
 \big\{(k^I-\la^I_1,k^F-\la^F_1);(k^I+u-\nu^I_1,k^F-\la^F_1)\big\}\ \ ,\cr
c\big\{(\la^I_1,\la^F_1);(\nu^I_1,\la^F_1)\big\}=&
 \delta_{\la^F_1,0}\big\{(\la^I_1,\la^F_1);(\nu^I_1,\la^F_1)\big\}\cr 
 \qquad\qquad&+(1-\delta_{\la^F_1,0})
 \big\{(k^I-\la^I_1,u-\la^F_1);(k^I+u-\nu^I_1,u-\la^F_1)\big\}\ \ . 
}}

The idea behind the proof is the following.
First, we notice that for the product ${\cal A}_{\lad\ladd}^{(\la^I,\la^F)}
{\cal A}_{\nud\nudd}^{(\nu^I,\la^F)}$ to be non-vanishing, 
$\la^F$ needs to be allowed by the A-type fusion of $\lad$ with $\ladd$ as
well as by the A-type fusion of $\nud$ with $\nudd$. However, due to
the branching conditions \cos, these prerequisites are equivalent, and we
conclude
\eqn\laFA{\eqalign{
 &{\cal A}_{\lad\ladd}^{(\la^I,\la^F)}
 {\cal A}_{\nud\nudd}^{(\nu^I,\la^F)}\neq0\ \ \ \ \ \ {\rm requires}\cr
 &\qquad\qquad\la^F_1=|\lad^F_1-\ladd^F_1|,\ 
  |\lad^F_1-\ladd^F_1|+2,\ ...,\ 
 k^F-|k^F-\lad^F_1-\ladd^F_1|\ \ .
}}
Similarly, for the B-type contributions we find
\eqn\laFB{\eqalign{
 &{\cal B}_{\lad\ladd}^{(\la^I,\la^F)}
 {\cal B}_{\nud\nudd}^{(\nu^I,\la^F)}\neq0\ \ \ \ \ \ {\rm requires}\cr
 &\qquad\qquad k^F-\la^F_1=|\lad^F_1-\ladd^F_1|+1,\ 
  |\lad^F_1-\ladd^F_1|+3\ ,\ ...,\ k^F-|k^F-\lad^F_1-\ladd^F_1|-1\ \ .
}}
Second, from \diag\ we read off the action of the combination $ac$ and its 
powers as
\eqn\acm{
(ac)^m\big\{(\la^I_1,\la^F_1);(\nu^I_1,\la^F_1)\big\}=
 \big\{(\la^I_1,\la^F_1-m);(\nu^I_1,\la^F_1-m)\big\}\ ,
 \ \ m\leq\la^F_1\neq0\ \ .
}

Using these results, we can mimic the pattern of \diag\ in the fusion
\strii, establishing that for any allowed $\la^F$ the summation
over integer parts ($\la^I$ and $\nu^I$) gives the same result as for
any other allowed $\la^F$:
\eqn\iso{\eqalign{
&\sum_{\la^I,\nu^I}{\cal A}_{\lad\ladd}^{(\la^I_1,k^F-|k^F-\lad^F_1-
 \ladd^F_1|)}{\cal A}_{\nud\nudd}^{(\nu^I_1,k^F-|k^F-\lad^F_1-\ladd^F_1|)}
  \cr &\qquad\times
 \big\{(\la^I_1,k^F-|k^F-\lad^F_1-\ladd^F_1|);
   (\nu^I_1,k^F-|k^F-\lad^F_1-\ladd^F_1|)\big\}\cr
&\qquad\cr
&=\sum_{\la^I,\nu^I}{\cal B}_{\lad\ladd}^{(\la^I_1,k^F-|k^F-\lad^F_1-
 \ladd^F_1|-1)}{\cal B}_{\nud\nudd}^{(\nu^I_1,k^F-|k^F-\lad^F_1-\ladd^F_1|-1)}
  \cr &\qquad\qquad\times
 \big\{(\la^I_1,k^F-|k^F-\lad^F_1-\ladd^F_1|-1);
   (\nu^I_1,k^F-|k^F-\lad^F_1-\ladd^F_1|-1)\big\}\cr
&\qquad\cr
&\vdots\cr
&\qquad\cr
&=\sum_{\la^I,\nu^I}{\cal B}_{\lad\ladd}^{(\la^I_1,|\lad^F_1-\ladd^F_1|+1)}
 {\cal B}_{\nud\nudd}^{(\nu^I_1,|\lad^F_1-\ladd^F_1|+1)}
 \big\{(\la^I_1,|\lad^F_1-\ladd^F_1|+1);
 (\nu^I_1,|\lad^F_1-\ladd^F_1|+1)\big\}\cr
&\qquad\cr
&=\sum_{\la^I,\nu^I}{\cal A}_{\lad\ladd}^{(\la^I_1,|\lad^F_1-\ladd^F_1|)}
 {\cal A}_{\nud\nudd}^{(\nu^I_1,|\lad^F_1-\ladd^F_1|)}
  \big\{(\la^I_1,|\lad^F_1-\ladd^F_1|);
 (\nu^I_1,|\lad^F_1-\ladd^F_1|)\big\}\ \ .
\cr &\qquad \cr
}}
{}From these identities it follows immediately that \strii\ reduces to
\eqn\dsum{\eqalign{
\sum_{\la^I,\nu^I}&{\cal A}_{\lad\ladd}^{(\la^I_1,\la^F_1)}
 {\cal A}_{\nud\nudd}^{(\nu^I_1,\la^F_1)}\big\{(\la^I_1,\la^F_1);(\nu^I_1
  ,\la^F_1) \big\}\ \ ,\cr
&{\rm for\ any\ of}\ \la^F_1=|\lad^F_1-\ladd^F_1|,\ |\lad^F_1-\ladd^F_1|+2,
 \ ...\ ,\ k^F-|k^F-\lad^F_1-\ladd^F_1|\ \ .
}}
Note that the fractional branching conditions in \cos\ still apply to the
expressions \iso\ and \dsum.
Of course, according to \iso\ we might as well have chosen to represent 
the final double-summation \dsum\ in terms of B-fusion coefficients, with
$\la^F$ subject to \laFB.
Though, this would only be possible provided
$|\lad^F_1-\ladd^F_1|+2\leq k^F-|k^F-\lad^F_1-\ladd^F_1|$, while
\dsum\ only requires $|\lad^F_1-\ladd^F_1|\leq k^F-|k^F-\lad^F_1-\ladd^F_1|$.

Now, substituting \ANN, the double summation \dsum\ becomes 
\eqn\dsumNN{
\sum_{\la^I,\nu^I}\, 
{\cal N}_{\dot\la^I\,\ddot\la^I}^{(k^I)\,\; \la^I}
 {\cal N}_{\dot\la^F\,\ddot\la^F}^{(k^F)\,\; \la^F}\ 
{\cal N}_{\dot\nu^I\ddot\nu^I}^{(k^I+u)\,\; \nu^I}
 {\cal N}_{\dot\la^F\,\ddot\la^F}^{(k^F)\,\; \la^F} 
\ \big\{(\la^I_1,\la^F_1);(\nu^I_1,\la^F_1)
 \big\}\ \ ,
}
with $\la^F$ subject to \laFA, in which case 
the two fusion coefficients 
at level $k^F$ are simply 1 and we reproduce the product form \fus. 
This concludes the proof that the ``supertrace'' reproduces  the 
minimal models fusion rules. 

The ``trace'' part of the proposal \str\ can actually 
be proved rather simply: the branching conditions on coset fields 
rule out the mixed combinations AAB and BAA. Here we have included
the A-label for the level-one integrable weights. 
Consider for instance the AAB situation. The fusion rules imply
the parity conditions 
\eqn\fuspar{\eqalign{
 (-1)^{\lad^I_1+\ladd^I_1+\la^I_1}=
 (-1)^{\mud^I_1+\mudd^I_1+\mu^I_1}=
 (-1)^{\nud^I_1+\nudd^I_1+\nu^I_1+k^I+u}=1&\ \ ,\cr
 (-1)^{\lad^F_1+\ladd^F_1+\la^F_1}=
 (-1)^{\nud^F_1+\nudd^F_1+\nu^F_1+u}=1&\ \ ,
}} 
while the branching conditions require 
\eqn\brapar{\eqalign{
 (-1)^{\lad^I_1+\mud^I_1+\nud^I_1+\lad^F_1}=
 (-1)^{\ladd^I_1+\mudd^I_1+\nudd^I_1+\ladd^F_1}=
 (-1)^{\la^I_1+\mu^I_1+\nu^I_1+\la^F_1}=1&\ \ ,\cr
 (-1)^{\lad^F_1+\nud^F_1}=
 (-1)^{\ladd^F_1+\nudd^F_1}=
 (-1)^{\la^F_1+\nu^F_1}=1&\ \ .
}} 
Firstly, substitute in the fusion condition 
$(-1)^{\mud^I_1+\mudd^I_1}=(-1)^{\mu^I_1}$
the integer branching conditions and compare the two sides using the fusion
conditions. From this we obtain $(-1)^{k^I+u}=1$. Secondly, 
a comparison of the fractional fusion and branching conditions yields 
$(-1)^u=1$.
It follows that both $k^I$ and $u$ must be even (and it is recalled
that $u=k^F+1$). However, that contradicts
$$k^F \quad {\rm odd}\quad \Ra \quad k^I \quad {\rm odd}\eqlabel\para$$
which is a consequence of \kIkF\ and the fact that $t$ and $u$ are 
relatively prime. The analysis of the case BAA is similar. 

Therefore, the proposal boils 
down to the introduction of minus signs in front of the BB-type 
contributions, cf. \str.

We should also stress that neglecting the intermediate weights at level 1 
is not an assumption as they can be reinserted everywhere using
the branching rule \branch\ and its dotted versions. 

Our proposal can also be stated by specifying what must replace the naive
factorisation in order to recover the correct coset fusions. From
\str, we find 
\eqn\superf{
 {\cal N}_{\lad\ladd}^{(k)\,\; \la}
 {\cal N}_{\mud\mudd}^{(1)\,\; \mu}
 {\cal N}_{\nud\nudd}^{(k+1)\,\; \nu} \ \ \rightarrow\ \ 
 {\cal A}_{\lad\ladd}^{(k)\,\; \la}
 {\cal N}_{\mud\mudd}^{(1)\,\; \mu}
 {\cal A}_{\nud\nudd}^{(k+1)\,\; \nu}-
 {\cal B}_{\lad\ladd}^{(k)\,\; \la}
 {\cal N}_{\mud\mudd}^{(1)\,\; \mu}
 {\cal B}_{\nud\nudd}^{(k+1)\,\; \nu}\ \ .
}
In the $\osp(1|2)$ analogy, this can be termed a
``super-factorisation''.  

\newsec{A coset truncation}

As an offshot of the supertrace analysis, we identify another way of 
recovering the coset fusion rules from the singular-vector decoupling 
method, namely via a truncation.

In \dsum, the fractional part can be chosen among any of the listed, possible 
values. Take for instance the upper limit. Then,  field identifications 
allow us to write
\dsum\ as
\eqn\dsumm{\eqalign{
 \sum_{\la^I,\nu^I}{\cal A}_{\lad\ladd}^{(\la^I_1,\la^F_1)}&
 {\cal A}_{\nud\nudd}^{(\nu^I_1,\la^F_1)}\big\{(\la^I_1,\la^F_1);(\nu^I_1
  ,\la^F_1) \big\}
 =\sum_{\la^I,\nu^I}{\cal A}_{(\lad^I_1,\lad^F_1),(\ladd^I_1,
  \ladd^F_1)}^{(\la^I_1,k^F-|k^F-\lad^F_1-\ladd^F_1|)}
 {\cal A}_{(\nud^I_1,\lad^F_1),(\nudd^I_1,\ladd^F_1)}^{(\nu^I_1,
  k^F-|k^F-\lad^F_1-\ladd^F_1|)}\cr
 &(aca)^{[(\lad^F_1+\ladd^F_1)/u]}
 \big\{(\la^I_1,k^F-|k^F-\lad^F_1-\ladd^F_1|);(\nu^I_1
  ,k^F-|k^F-\lad^F_1-\ladd^F_1|) \big\}\ \ ,
}}
This clearly resembles the result \cosfi\ obtained from the Verlinde
formula.\foot{The precise form of the formula depends upon the specific
choice made for $\la^F_1$.  For instance, with
$$
 \la^F_1=k^F-|k^F-\lad^F_1-\ladd^F_1|-2n\ \ \ \ {\rm with}\ \ 2n=0,2,...,
 k^F-|k^F-\lad^F_1-\ladd^F_1|-|\lad^F_1-\ladd^F_1|\ \ ,
\eq
$$
the part $(aca)^{[(\lad^F_1+\ladd^F_1)/u]}
\{\cdots\}$ would be replaced by
$$
(aca)^{[(\lad^F_1+\ladd^F_1)/u]}(ca)^{2n}
 \big\{(\la^I_1,k^F-|k^F-\lad^F_1-\ladd^F_1|-2n);(\nu^I_1
  ,k^F-|k^F-\lad^F_1-\ladd^F_1|-2n) \big\}
$$
The upper limit ($n=0$) is singled out as the more natural choice when 
comparing with the Verlinde formula as it is the only case where
there are no required field identifications when
$\lad^F_1+\ladd^F_1<u$. The conclusion concerning associativity is
general in that it holds for all values of $n$.}
This is an example of a truncation
of the singular-vector  decoupling fusions that is sufficient to reproduce the
correct coset fusions (see also Section 10).  

It should be stressed, however, that truncating a set of fusion rules known to
satisfy fundamental, albeit delicate, compatibility requirements such as
associativity, is bound to put these  properties in peril. Indeed, the 
truncated fusions presented above
are not associative. Technically, this is a reflection of the non-identity
$$
\delta_{\nu^F_1,k^F-|\mu^F_1-
|k^F-\la^F_1-\rho^F_1||}\neq\delta_{\nu^F_1,k^F-|\la^F_1-
|k^F-\mu^F_1-\rho^F_1||}
$$ 
This indicates the restricted usefulness of such a truncation. 

Nevertheless, this simple analysis provides
a second prescription for recovering the coset fusions: they may be
expressed in terms of a (non-associative) truncation  of the singular-vector
decoupling fusions. The truncation amounts to considering only type A 
fusions and
moreover,  only with a fixed value of the  fractional part of the evaluated
weights. The truncation can in most cases also be expressed in terms of type B
fusions (cf. discussion following \dsum).

\newsec{WZW fusion rules: relating the two methods} 

In this section, we discuss a correspondence between  
other truncated versions of the fusion obtained by singular-vector
decoupling, and the fusion computed using the Verlinde formula. 
This correspondence relies on the notion of highest- vs lowest-weight 
condition as applied to a triple product, and is motivated by the note
added to [\AY]. 

The decoupling method is based on a study of three-point chiral blocks
of the form $\langle\nu|\phi_\mu|\la\rangle$.  
In that context, it makes sense to discuss the possible representations
carried by the middle field and use those to characterise the associated
fusions. Thus, we have the following highest- and lowest-weight
conditions (see [\AY] and [\PRYR])
$$
\eqalignD{
&{\rm hwc}:\qquad\quad &{1\over 2}(\la_1+\mu_1-\nu_1)\in\Z_{\geq}\ \ ,\cr
&{\rm lwc}:\qquad &{1\over 2}(-\la_1+\mu_1+\nu_1)\in\Z_{\geq}\ \ .\cr}
\eqlabel\hlwc
$$
A fusion satisfying neither of these two possibilities belongs
to the so-called continuous series.
Note that the resulting fusion rules (\ayf) are symmetric despite
the asymmetric starting point $\langle\nu|\phi_\mu|\la\rangle$
and its associated assignments (\hlwc).

We shall demonstrate that B-fusions belong to the continuous series.
First we rule out the possibility that a B-fusion can satisfy the
highest-weight condition. In terms of integer and fractional
parts, this condition reads
$$ 
{1\over 2}(\la^I_1+\mu^I_1-\nu^I_1)- {1\over 2}(k+2)(\la^F_1+\mu^F_1-\nu^F_1)
\in\Z_{\geq}\ \ .\eq
$$
It follows immediately that
$$
\la^F_1+\mu^F_1-\nu^F_1= 0 \quad {\rm or}\quad u\ \ .\eqlabel\hwsco
$$
A comparison with the fractional B-fusion yields that $k^F$ is odd
in the first case ($\la^F_1+\mu^F_1-\nu^F_1=0$)
while a contradiction is obtained in the second ($\la^F_1+\mu^F_1-\nu^F_1=u$).
To rule out the first case, we consider the integer part from which it
follows that $k^I$ is even, in contradiction with (\para).

Second, the lowest-weight condition reads
$$ 
{1\over 2}(-\la^I_1+\mu^I_1+\nu^I_1)
 -{1\over 2}(k+2)(-\la^F_1+\mu^F_1+\nu^F_1)\in\Z_{\geq}\ \ .\eq
$$
and as before, there are two possibilities for the fractional part
$$
-\la^F_1+\mu^F_1+\nu^F_1=0\quad {\rm or}\quad u\ \ .\eqlabel\lwsco
$$
Again, the first case is ruled out by (\para), while the second case
contradicts the fractional fusion rule itself.

A highest- or lowest-weight fusion can thus only be of type A.  
But not all type-A fusions respect the highest- or lowest-weight 
conditions. 
Again, the highest-weight condition (\hwsco) is compatible with the 
A-type fusion rules only if $\nu^F_1=\la^F_1+\mu^F_1$
(in which case $\la^F_1+\mu^F_1<u$), while the lowest-weight
condition (\lwsco) is compatible with the A-type fusion rules only if
$\nu^F_1=\la^F_1-\mu^F_1$ (in which case $\mu^F_1\leq\la^F_1$).
It is seen that the highest- and lowest-weight fusions correspond
to the upper and lower bounds, respectively, of the
fractional part of the type-A fusion. The integer
parts are not constrained further than by the A-fusion rule.

We note that a fusion can satisfy both the highest- and the lowest-weight
conditions. That happens precisely when $\mu^F_1=0$ (and $\la^F_1=\nu^F_1$), 
in which case the integer part of the type-A fusion ensures that both the
conditions in (\hlwc) are satisfied. We may therefore group the highest-
and lowest-weight fusions into two disjoint sets: the highest-weight
fusions and the purely lowest-weight fusions, respectively.
The latter are thus characterised by $\mu^F_1=\la^F_1-\nu^F_1>0$.

It is natural to symmetrise the notion of lowest-weight fusion:
$$
 {\rm slwc}:\ \ \ \ {1\over 2}(-\la_1+\mu_1+\nu_1)\in\Z_\geq\ \ \
 {\rm or}\ \ \ {1\over 2}(\la_1-\mu_1+\nu_1)\in\Z_\geq\ \ ,\eqlabel\slwc
$$
and denote the two cases lowest-weight fusions with respect to $\mu$
or $\la$, respectively. The latter case corresponds to
studying the three-point chiral block 
$\langle\nu|\phi_\la|\mu\rangle$. It follows immediately that purely 
$\la$-lowest-weight fusions are of type A and characterised by 
$\nu^F_1=-\la^F_1+\mu^F_1$ (in which case $\la^F_1\leq\mu^F_1$).
A symmetrised lowest-weight fusion (\slwc) is then 
characterised by $\nu^F_1=|\la^F_1-\mu^F_1|$, while an associated purely
lowest-weight fusion respects in addition $\la^F_1,\mu^F_1\geq1$.
Hence, the symmetrised lowest-weight condition corresponds precisely to the
lower bound of the fractional part of the type-A fusion.

We now construct a precise correspondence between the set of highest- or
lowest-weight fusions from singular-vector 
decouplings (fusions known to be of type A) and the Verlinde formula.
Highest-weight fusions  
correspond to positive Verlinde fusions, whereas purely lowest-weight
fusions correspond to negative Verlinde fusions.
In the first case, the correspondence is obvious as it is
precisely the condition $\la^F_1+\mu^F_1<u$ that ensures the positivity
in the Verlinde formula. The integer parts also match.

The correspondence between lowest-weight fusions and negative Verlinde
coefficients is less direct. It is 
established through simple manipulations of the Verlinde fusion
rules. The ``negative part'' of the Verlinde fusion rule (\frafus) can 
be written as (with the understanding that 
$\tilde\la^F_1+\tilde\mu^F_1-u\geq 0$)
$$
(\tilde\la_1^I,\tilde\la_1^F)\times (\tilde\mu^I_1,\tilde\mu_1^F)\  \supset
\ -\sum_{\tilde\nu^I_1\in \delta_{k^I}
} ~(k^I-\tilde\nu^I_1,\tilde\la^F_1+\tilde\mu^F_1-u )
 \ \ .\eqlabel\frafusa
$$
where the tildes have been introduced to avoid confusion later. The 
set $\delta_{k^I}$ is defined as in (\sett), in terms of 
$\tilde\la,\ \tilde\mu$ and $\tilde\nu$.
Replacing $\tilde\nu^I_1$ by
$k^I-\tilde\nu^I_1$ modifies the range $\delta_{k^I}$ to
$$
 \delta'_{k^I}= \{\tilde\nu^I_1\ |\ \tilde\la^I_1+
\tilde\mu^I_1-\tilde\nu^I_1+k^I\in2\Z_{\geq}
  ;\quad
 |k^I-
\tilde\la^I_1-\tilde\mu^I_1|\leq \tilde\nu^I_1 \leq 
k^I-|\tilde\la^I-\tilde\mu^I_1|\}\ \ .\eq
$$

On the other hand, recall that a purely $\mu$-lowest-weight 
fusion respects $1\leq\mu^F_1\leq\la^F_1$.
This means that the sum of the fractional parts of $\la$ and $c\mu$ is
$\la^F_1+(u-\mu^F_1)\geq u$.  This corresponds to the condition for a negative
Verlinde fusion. This suggests that instead of (\frafusa), we should consider
$$
 (\la_1^I,\la^F_1)\times c(\mu_1^I,\mu_1^F)\  \supset\  -\ 
\sum_{\nu^I_1\in \delta^{(c)}_{k^I}} ~(\nu^I_1,\la^F_1-\mu^F_1)
  \ \ .\eqlabel\frafusa
$$
where
$$
 \delta^{(c)}_{k^I}= \{\nu^I_1\ |\ \la^I_1-\mu^I_1-\nu^I_1+2k^I
  \in2\Z_{\geq};\quad
 |\la^I_1-\mu^I_1|\leq \nu^I_1 \leq k^I-|k^I-\la^I-\mu^I_1|\}\ \ .\eq
$$
The conditions in $\delta^{(c)}_{k^I}$
imply that $\la^I_1+\mu^I_1+\nu^I_1\in2\Z_{\geq}$; since $\nu_1^F=
\la^F_1-\mu^F_1$,  we have all the characteristics of a purely
$\mu$-lowest-weight fusion (see (\hlwc)). The analogous correspondence between
$\la$-lowest-weight fusions and the negative Verlinde fusions is based on 
$c\la\times\mu$ instead of $\la\times c\mu$.

The correspondence between highest-weight fusions and the positive Verlinde
fusions is obviously one-to-one. This is not the case for the correspondence
between one of the two types of purely lowest-weight fusions and the negative
Verlinde fusions. 
A natural way of defining a one-to-one map between the negative Verlinde
fusions and the purely lowest-weight fusions is to group the latter in 
equivalence classes:
$$
 [\la,\mu]\ \sim\ [c\la,c\mu]\ \ .\eqlabel\equiv
$$
Note that $c$ always maps a pure and symmetrised lowest-weight onto
a pure and symmetrised lowest-weight (while that is not a property of $a$).
We also observe that when $\la\neq\mu$, (\equiv) relates a
$\mu$-lowest-weight fusion to a $\la$-lowest-weight fusion.
When $\la=\mu$, both fusions are lowest-weight fusions with respect
to both $\mu$ and $\la$.
The sought one-to-one map is between these equivalence classes of
symmetrised, purely lowest-weight fusions and negative Verlinde 
fusions.\foot{
Alternatively, one may work with the $\mu$-lowest-weight fusions alone,
for example. In that case, one must choose a set of representatives
for the equivalence classes. When $\la\neq\mu$, the choice is unique,
while for $\la=\mu$ there are several. Natural choices are
either $\la^F_1+\mu^F_1\leq u$ or $\la^F_1+\mu^F_1\geq u$.
Having made such a choice, the correspondence is one-to-one
between the accordingly reduced (or truncated) purely 
$\mu$-lowest-weight fusions and the negative Verlinde fusions.}

We have thus established a one-to-one correspondence between 
different truncations  
of the A-type decoupling fusions and the full set of Verlinde
fusions. This clarifies and corrects the corresponding 
statements made in a note added to [\AY].

Finally, we want to comment on yet another truncation of the WZW fusions
that reproduces the coset fusions. It combines the notion of
highest- and lowest-weight fusions with the result of the previous
section. 

First we observe that for each possible value of the sum $\lad^F_1+\ladd^F_1$,
we may choose a representative of \dsum\ --  see \dsumm\ and the subsequent 
footnote. A particular set of choices is reflected in the expression
\eqn\dsumu{\eqalign{
 \left(1-[(\lad^F_1+\ladd^F_1)/u]\right)&\sum_{\la^I,\nu^I}
  {\cal A}_{(\lad^I_1,\lad^F_1),(\ladd^I_1,
  \ladd^F_1)}^{(\la^I_1,\lad^F_1+\ladd^F_1)}
 {\cal A}_{(\nud^I_1,\lad^F_1),(\nudd^I_1,\ladd^F_1)}^{(\nu^I_1,
  \lad^F_1+\ladd^F_1)}
 \big\{(\la^I_1,\lad^F_1+\ladd^F_1);(\nu^I_1,\lad^F_1+\ladd^F_1) \big\}\cr
 +[(\lad^F_1+\ladd^F_1)/u]&
  \sum_{\la^I,\nu^I}{\cal A}_{(\lad^I_1,\lad^F_1),(\ladd^I_1,
  \ladd^F_1)}^{(\la^I_1,|\lad^F_1-\ladd^F_1|)}
 {\cal A}_{(\nud^I_1,\lad^F_1),(\nudd^I_1,\ladd^F_1)}^{(\nu^I_1,
  |\lad^F_1-\ladd^F_1|)}\big\{(\la^I_1,|\lad^F_1-\ladd^F_1|);(\nu^I_1
  ,|\lad^F_1-\ladd^F_1|) \big\}\ \ ,
}}
Note that due to the prefactor of the first term, 
$\la^F_1=k^F-|k^F-\lad^F_1-\ladd^F_1|$ is reduced to $\lad^F_1+\ladd^F_1$.

Let H and L stand for highest-weight fusions and purely
lowest-weight fusions, respectively, while L$_\geq$ represents an L
subject to $\lad^F_1+\ladd^F_1\geq u$. It is emphasised that L$_\geq$
simply corresponds to choosing a particular equivalence class
representative. From \dsumu\ we thus see, that one need only truncate
the full set of triple-field products obtained from the separate WZW
fusions to those of type HHH or L$_\geq$HL$_\geq$, 
where the three labels refer to the representations at respective levels 
$k$, $1$ and $k+1$ (the middle field is associated to a highest-weight 
fusion H, as L$_\geq$ denote a {\it purely} lowest-weight fusion,
or more precisely, an equivalence class representative).
It is also natural to end up 
with a requirement specifying that in coset fields, highest- 
(respectively, lowest-) weight fusions are to be combined with highest- 
(respectively, lowest-) weight fusions.
Note, however, that only a subset of the lowest-weight fusions are used.

Incidentally, restricting to the set of symmetrised
(but not necessarily purely) lowest-weight fusions, likewise reproduces the 
coset fusions. This situation occurs when choosing 
$\la^F_1=|\lad^F_1-\ladd^F_1|$ in \dsum.

\newsec{Conclusion}

Our results aim to fill an obvious gap in the coset description of
non-unitary rational CFTs. Such coset CFTs involve WZW models at
fractional, admissible levels. Their fusion rules are
ambiguous: Verlinde fusion differs from
(singular-vector) decoupling fusion. Only the Verlinde fusion rules
were known to be consistent with the unambiguous coset fusion, in the
simplest case of diagonal cosets 
involving one integer level. Here we have shown how
the decoupling fusion can yield the correct fusion in the Virasoro
minimal models, described by the diagonal coset (\coco). We restricted
to this coset for simplicity, and because decoupling fusion has only
been worked out completely in the $\su(2)$ case.\foot{Complete results
for certain levels are now known for $\su(3)$ [\ref{P. Furlan,
A.Ch. Ganchev and V.B. Petkova, 
{ Nucl. Phys}. {\bf B518} (1998) 645, Commun. Math. Phys. {\bf 202} 
(1999) 701; A.Ch. Ganchev, V.B. Petkova and G. Watts, { Nucl. Phys}. {\bf
B571} (2000) 457. }].}

We should emphasise again that WZW models at fractional levels may not be
consistent, except in coset theories. After all, the Verlinde fusion
coefficients are sometimes negative integers. These negative signs do
cancel in the diagonal coset CFTs, however. The key input is the
factorisation \cosf, a consequence of the factorisation of the
corresponding coset modular $S$ matrix.  

We find here that although the decoupling fusion coefficients are
non-negative, they are too big for the cosets. That is, they lead to coset
fusion coefficients that are too large, when factorisation is
assumed. By introducing minus signs at the coset level, we found a
prescription that led to the correct coset fusions. 

These minus signs cause cancellations, and so are
equivalent to a truncation. We also found direct descriptions for
such truncations. Particular attention was paid to a truncation
motivated by the notion of highest and lowest weights in the
derivation of the singular-vector decoupling fusions. 

The first prescription is partly motivated by its interpretation as a
``super-trace'' or ``super-factorisation'', in the $\osp(1|2)$ analogy 
found previously. The second is natural in the coset
description, since it pairs highest weights with highest weights, and
lowest weights with lowest weights. 

Of course, a deeper motivation should be found for our recipes. Perhaps
the relation between the singular vectors of algebra and subalgebra
can provide it. We leave that to future work. 

\vskip 1truecm
\noindent{\it Acknowledgements}

Part of this work was done during the workshop {\it Quantum integrability} 
at the CRM, Universit\'e de Montr\'eal. We thank the CRM for 
its generous hospitality.

\vfill\eject
\centerline{\bf REFERENCES}
\vskip 1cm
\immediate\closeout\refs \vskip 0.5cm
  \message{References}\input references
\vfill\eject

\end